\documentclass[11pt,a4paper]{article}
\usepackage{jheppub}
\usepackage{amsfonts}
\usepackage{amssymb}
\usepackage{textcomp}
\everymath{\displaystyle}

\title{ Phase structure of fuzzy black holes}

\author[a]{S. Digal,} 

\author[a]{T. R. Govindarajan,}

\author[b]{Kumar S. Gupta,}

\author[c]{X. Martin}

\affiliation[a]{Institute of Mathematical Sciences,\\
 Chennai, 600113, India}

\affiliation[b]{Theory Division, Saha Institute of Nuclear Physics,\\
1/AF Bidhannagar, Calcutta 700064, India}

\affiliation[c]{LMPT, UFR Sciences et Techniques, Universite de Tours,\\ 
Parc de Grandmont, 37200 TOURS, France}

\emailAdd{digal@imsc.res.in}
\emailAdd{trg@imsc.res.in}
\emailAdd{kumars.gupta@saha.ac.in}
\emailAdd{xavier@lmpt.univ-tours.fr}

\abstract{Noncommutative deformations of the BTZ black holes are described
by noncommutative cylinders. We study the scalar fields in this background. The spectrum is studied analytically 
and through numerical simulations we establish the existence of novel `stripe phases'. These are different
from stripes on Moyal spaces and stable due to topological obstruction.}


\begin{document}

\maketitle

\flushbottom

\def\be{\begin{equation}}
\def\ee{\end{equation}}
\def\beq{\begin{eqnarray}}
\def\eeq{\end{eqnarray}}
\def\bn{\begin{eqnarray*}}
\def\en{\end{eqnarray*}}
\def\slas{\!\!\!/}
\def\BI{{\rm 1\!l}}
\def\P{\Phi}
\def\p{\phi}
\def\w{\omega}
\def\W{\Omega}
\def\O{{\cal{O}}}
\def\a{\alpha}
\def\b{\beta}
\def\s{\sigma}
\def\S{\Sigma}
\def\d{\delta}
\def\D{\Delta}
\def\g{\gamma}
\def\t{\theta}
\def\T{\Theta}
\def\G{\Gamma}
\def\z{\zeta}
\def\Z{\Psi}
\def\pd{\partial}
\def\e{\epsilon}
\def\n{\eta}
\def\m{\mu}
\def\r{\rho}
\def\t{\theta}
\def\R{\Rho}
\def\bra{{\langle}}
\def\ket{{\rangle}}
\def\bp{{\bf p}}
\def\bq{{\bf q}}
\def\bk{\bf k}
\def\br{{\bf r}}
\def\bx{{\bf x}}
\def\by{{\bf y}}
\def\l{\lambda}
\def\L{\Lambda}
\def\cL{{\cal{L}}}
\def\cH{{\cal{H}}}
\def\cP{{\cal{P}}}
\def\cU{{\cal{U}}}
\def\cN{{\cal{N}}}
\def\cT{{\cal{T}}}
\def\cC{{\cal{C}}}
\def\cD{{\cal{D}}}
\def\cJ{{\cal{J}}}
\def\cK{{\cal{K}}}
\def\cM{{\cal{M}}}
\def\cA{{\cal{A}}}
\def\cB{{\cal{B}}}
\def\cO{{\cal{O}}}
\def\cR{{\cal{R}}}
\def\cG{{\cal{G}}}
\def\cS{{\cal{S}}}
\def\cF{{\cal{F}}}
\def\cI{{\cal{I}}}
\def\cZ{{\cal{Z}}}
\def\la{\langle}
\def\ra{\rangle}

\newcommand{\diff}{\mathrm{d}}

\section{Introduction}

General relativity and quantum mechanics together imply that space-time structure at the Planck scale is described by noncommutative geometry \cite{dop2}. There have been various attempts to study gravity theories within the noncommutative framework \cite{wess1,seckin}. This has led to a Hopf algebraic description of noncommutative black holes \cite{brian1,schupp} and FRW cosmologies \cite{ohl}. A large class of such black hole solutions, including the noncommutative BTZ \cite{btz1,btz2} and Kerr black holes, exhibits an universal feature where the Hopf algebra is described by a noncommutative cylinder \cite{petergrosse}, which belongs to the general class of the $\kappa$-Minkowski algebras \cite{L1,L4,M1,M3}. For the purpose of this paper, we shall take the noncommutative cylinder and the associated algebra as a prototype for noncommutative black holes. 

The study of quantum field theories in the background of black holes has led to the discovery of interesting features associated with the underlying geometry, such as the Hawking radiation and black hole entropy. In the noncommutative case, the black hole geometry is replaced with the algebra defined by the noncommutative cylinder. In order to probe the features of a noncommutative black hole, it is useful to analyze the behaviour of a quantum field coupled to the noncommutative cylinder algebra. Scalar field theories have been extensively studied on $\kappa$-Minkowski spaces \cite{luk2,luk3,G1,G2,ksms}, which has led to twisted statistics and deformed oscillator algebra for the quantum field \cite{G1,G2,dice2010}. Theories on the noncommutative cylinder lead to quantization of the time operator \cite{C2,paulo}

The quantum field theories defined on noncommutative spaces are highly nonlocal and in order to gain further insight into their behaviour, it is essential to simulate their behaviour through numerical analysis. To this end, it is necessary to approximate the infinite dimensional noncommutative cylinder algebra with a suitable truncated finite dimensional matrix algebra, belonging to the general class of 
fuzzy spaces \cite{madore,balbook,hoppe}. 

Field theories on noncommutative geometries are inherently as
mentioned earlier non-local leading to mixing of infrared and
ultraviolet scales. This, in turn, is responsible for new ground
states with spatially varying condensates. Many non-perturbative
studies have established that noncommutative spaces, such as the
Groenewold-Moyal plane and fuzzy spheres, allow for the formation of
stable non-uniform condensates as ground states.  Exploring the
implications of the non-local nature of field theories is very important
in many areas of quantum physics. \cite{pinzul,gubser}

Since different phases are intimately connected with spontaneous
symmetry breaking (SSB), the role of symmetries in noncommutative
geometries themselves is subtle.  This issue is important in 2D
because the Coleman-Mermin-Wagner (CMW) theorem states that there can
be no SSB of continuous symmetry on 2-dimensional commutative
spaces. There is no obvious generalization of the CMW theorem for
non-commutative spaces, since the theorem relies strongly on the
locality of interactions.  Noncommutative spaces admit non-uniform
solutions (in the mean field) and one can ask the question what
happens to the stability of these configurations. Non-uniform
condensates naturally have a infra-red cut-off for the
fluctuations. This cut-off softens the otherwise divergent
contributions of the Goldstone modes\cite{xavier,denjoe,digal1,digal2,
  bietenholz,ambjorn,panero,medina}.

This paper is organized as follows. In Sec. 2 we give a brief
introduction to fuzzy black holes motivated by an earlier analysis of
the noncommutative deformation of Banados-Teitelboim-Zanelli (BTZ)
black holes. In Sec. 3 we set up the algebra describing the
noncommutative cylinder which is suitable for numerical
simulations. In Secs. 4 and 5, we provide the action for the scalar
field and the spectrum of the Laplacian on the noncommutative cylinder
respectively.  Sec. 6 exhibits the phase structure and the novel
stripe phases which are generic to fuzzy spaces. We exhibit the
crucial differences of these different phases and analyze our results
in Sec. 7.

\section{Fuzzy black holes}

We briefly summarize the essential features of a noncommutative black hole which is useful for our analysis. In the commutative case, 
a non-extremal BTZ black hole is described in terms of the coordinates $(r, \phi, t)$ and is given by the metric \cite{btz1,btz2}
\be 
ds^2= \biggl( M - \frac {r^2}{\ell^2} - \frac{J^2}{4r^2}\biggr)
dt^2  + \biggl( -M + \frac {r^2}{\ell^2} +
\frac{J^2}{4r^2}\biggr)^{-1} dr^2+ r^2 \biggl(d\phi - \frac J{2r^2} dt\biggr)^2
\;,
\label{btzmtrc}
\ee
where $0\le r<\infty\;, \;-\infty < t<\infty\;, \;0\le \phi<2\pi\;,$ $M$ and
 $J$ are respectively the mass and spin of the black hole, and  $\Lambda= -1/\ell^2$ is the 
cosmological constant. In the non-extremal case, the two distinct horizons $ r_{\pm}$ are given by
\be 
r_\pm^2 =\frac {M\ell^2}2 \biggl\{ 1\pm \bigg[ 1 - \biggl(\frac
J{M\ell}\biggr)^2\biggr]^{\frac 12} \biggr\}. \;
\label{rpm}
\ee

An alternative way to obtain the geometry of the BTZ black hole is to
quotient the manifold $AdS_3$ or $SL(2,R)$ by a discrete subgroup of
its isometry. The noncommutative BTZ black hole is then obtained by a
deformation of $AdS_3$ or $SL(2,R)$ which respects the quotienting
\cite{brian1}.  In the noncommutative theory, the coordinates $r$,
$\phi$ and $t$ are replaced by the corresponding operators $\hat r$,
$\hat \phi$ and $ Z $ respectively, that satisfy the algebra
\be 
[\hat{t}, e^{i\hat \phi}] = \alpha e^{i\hat \phi}\qquad [\hat r,\hat t] =  
[\hat r,e^{i\hat \phi}] =0\;,
\label{qntmalg} 
\ee
where the constant $\alpha$ is proportional to $\ell^3/(r_+^2-r_-^2)$. We shall henceforth refer to (\ref{qntmalg}) 
as the noncommutative cylinder algebra. Furthermore, $\hat t$  denoting the operator corresponding the the axis of the cylinder, it will be therefore identified as the operator $Z$ also in the following sections.

It may be noted that the operator $\hat r$ is in the center of the algebra (\ref{qntmalg}). In addition, 
it can be shown easily that $e^{-2\pi i\hat t/\alpha}$ belongs to the center of (\ref{qntmalg}) as well. 
Hence, in any irreducible representation of (\ref{qntmalg}), the element $e^{-2\pi i\hat t/\alpha}$ is proportional 
to the identity,
\be
e^{-2\pi i\hat t/\alpha} = e^{i\gamma}\BI,
\label{central}
\ee
where $\gamma \in R$ mod $(2 \pi)$. Eqn. (\ref{central}) implies that in any irreducible 
representation of (\ref{qntmalg}), the spectrum of the time operator $\hat{t}$, or $Z$, is quantized \cite{petergrosse,C2,paulo} and is given by
\be 
{\rm {spec}}~ {\hat t} = n \alpha
-\frac {\gamma\alpha}{2\pi}\;,\;\;n\in
{\mathbb{Z}}.
\label{dsctsptmt}
\ee
In what follows we shall set $\gamma = 0$ without loss of generality.

The noncommutative cylinder algebra (\ref{qntmalg}) belongs to a special class of the $\kappa$-Minkowski 
algebra and it appears in the description of noncommutative Kerr black holes \cite{schupp} and FRW cosmologies \cite{ohl}. 
We shall henceforth consider (\ref{qntmalg}) as a prototype of the noncommutative black hole.

\section{Noncommutative Cylinder Algebra}

The NC cylinder is defined by the relation
\be
\left[~Z,~e^{i \phi}~\right]~=~\a e^{i \phi}
\label{angle}
\ee
where $Z$ is hermitian and $e^{i \phi}$ is unitary. As mentioned earlier, the operator $Z$ corresponds to the axis of the cylinder, and therefore to the time operator $\hat t$ of the black hole. Since we are interested in simulations, we have to discretize the above NC cylinder.

In the rest of this paper, we will work with $\alpha=1$ without loss
of generality, since the simple scaling $Z\rightarrow Z/\alpha$ can
scale $\alpha$ away in the commutation relation (\ref{angle}).

For this purpose consider the spin $J$ irreducible representation (IRR) of the $SU(2)$ Lie algebra, given by
\be
[X_+,X_-]~=~2X_3, ~~[X_\pm, X_3]~=~\mp X_\pm. \label{CR}
\ee  
Since the operator $Z$ generates rotations around the axis of the cylinder, it can be identified with $X_3$.
But when we use the finite dimensional representations 
of $SU(2)$ we cannot implement (\ref{angle}) with unitarity for $e^{i \phi}$.
For this purpose, we decompose $X_+$ as product of a unitary and a Hermitian  
operator as given by
\be
X_+~=~e^{i\phi}~R \label{defRe}
\ee
In (\ref{defRe}), $e^{i\phi}$ is unitary, and $R$ is a positive Hermitian,
necessarily singular, matrix which commutes with $Z$ (and is thus diagonal). Using (\ref{CR}) and the fact that $R$ commutes with $Z$, we have  
\be
[Z, X_+]~=~[Z, e^{i \phi}]~R~=~ e^{i \phi}~R \label{ZEcom}
\ee
Since $R$ is singular, it can not be inverted. However a partial inverse
$\tilde{R}$ can be found such that $R \tilde{R}~=~P$, the projector such that
$1-P$ projects on the kernel of $R$. Thus we get
\be 
~[Z, e^{i \phi}]~P=~ e^{i \phi}~P\label{Pcr}
\ee

To find a representation for $R$ and $e^{i \phi}$, we can look at 
\beq 
R^2=X_-X_+=\vec{L}^2-L_3(L_3+1)
\eeq
which commutes with $Z=L_3$. Remember that in the usual representation of angular momentum, 
$L_3|l,m>=m|l,m>$ with $|m|\leq l$. Shifting the indices from $0$ to
$2l+1=2J$, we have $j=m+l+1$, leading to
\begin{eqnarray*}
X_-X_+|l,m> & = & [l(l+1)-m(m+1)]|l,m>=[(l+1/2)^2-(m+1/2)^2]|l,m>\\
X_-X_+|J,j> & = & [J^2-(J-j)^2]|J,j>=j(2J-j)|J,j> \end{eqnarray*}
There is only one hermitian positive solution to this equation which takes the
form  
\be 
R_{ij}~=~\sqrt{i(2J-i)}~\d_{i,j}
\ee
which is diagonal as expected, and whose null space is along the top
state $|J,2J>$. As a result, $P=1-|J,2J><J,2J|$, and  
$$\tilde{R}=\sum_{i=1}^{2J-1} [i(2J-i)]^{-1/2}\,|J,i><J,i|$$
where the sum stops at $i=2J-1$ so that there is a zero in the last position on
the diagonal.

It is now possible to deduce the first $2J-1$ lines of the unitary matrix
$e^{i \phi}$ from (\ref{defRe}): 
$$\left.\begin{array}{rcl} X_+|J,j> & = & \sqrt{j(2J-j)}|J,j+1>\\
X_+|J,j> & = & e^{i\phi}~R|J,j>=\sqrt{j(2J-j)}e^{i\phi}~|J,j>
\end{array}\right\} \Rightarrow e^{i\phi}~|J,j>=|J,j+1>,\ j<2J.$$
Eq. (\ref{defRe}) yields no equation for the last column which is
instead determined from its unitarity. The columns 
$1,\cdots,2J-1$ of $e^{i \phi}$, given by $|J,j+1>$, form an orthonormal set,
as expected for a unitary matrix. Then the last
column will be a vector orthogonal to all these vectors and thus can only be
proportional to $|J,1>$. After normalization that still leaves a $U(1)$
freedom so that
\be
(e^{i \phi})_{ij}~=~ \d_{i,j+1}~+~ e^{i\b}~\d_{i,1}\d_{j,2J}
\label{newangle}
\ee
where $\b$ can be any real number. For $\b=0$, $e^{i \phi}$ is just a
circular permutation of length $2J$.

\section{The Action on the fuzzy cylinder}
We will first construct an action for a hermitian scalar field $\Phi$.
Define $ \widetilde{Tr} O~=~Tr(P O P) $. 
This trace $ \widetilde{Tr} $ is equivalent to integrating over 
the whole cylinder in the continuum limit. We also need the derivatives $ \pd_\phi $ and $ \pd_Z $. They are:
\beq
\pd_\phi \Phi~&=&~[Z,\Phi] \label{Zgen} \\
\pd_Z\Phi~&=&~ e^{-i\phi}[e^{i\phi},\Phi]
\eeq
Then, apart from $J$-dependent normalization factors, a naive form of the
action can be chosen as:
\be 
S~=~\widetilde{Tr}\left(|~[Z,\Phi]~|^2 ~+~ |~e^{-i\phi}[e^{i\phi},\Phi]~|^2~+~
V(\Phi)\right) \label{Sgen}
\ee
where $V(\Phi)$ is the potential which can be taken to be of the form,
\be
V(\Phi)=\mu\Phi^2+c\Phi^4 \label{potential}
\ee
for a hermitian field $\Phi$. 

This action has a problem of instability. The source of this comes from
$\widetilde{Tr}(\Phi^4)=Tr((P\Phi)\Phi^2(\Phi P))$ which cannot contain any
quartic (nor cubic) term for the variable $\Phi_{2J\,2J}$. 
This makes the theory unstable with respect to this variable.
The simplest cure is to insist that this term is not a degree of freedom
of the theory and constrain it to zero. To keep the set of fields an algebra, we choose to also set to zero the last row $\Phi_{2J\,i}$ and column $\Phi_{i\,2J}$ of the field. As a result the hermitian field
$\Phi$ now only has $(2J-1)^2$ degrees of freedom, and $\Phi=P\Phi P$.

\subsection{Dimensional reduction}
With this new choice of the field  the action becomes 
\beq
S & = & Tr\left(~P~|~[Z,P\Phi P]~|^2P~+~P~|~e^{-i\phi}[e^{i\phi},P\Phi P]~|
^2~P+~V(\Phi)\right) \\ 
& = & Tr\left(|~[PZP,\Phi]~|^2~+|~[Pe^{i\phi}P,\Phi]~|^2+~V(\Phi)\right)
\eeq
which can be rewritten simply as the action for a hermitian matrix in a
$(2J-1)\times(2J-1)$ matrix algebra of reduced dimension:
\be S~=~Tr_{J'}\left(|~[\tilde{Z},\Phi]~|^2~+|~[\widetilde{e^{i\phi}},
\Phi]~|^2+~V(\Phi)\right) \label{reducedS}
\ee
where $J'=J-1/2$ is the reduced angular momentum, while $\widetilde{e^{i
\phi}}$ and $\tilde{Z}$ are the matrices obtained from $e^{i\phi}$ and $Z$
by removing the last line and column. For $\widetilde{e^{i
\phi}}$, it is equivalent to setting $e^{i\b}\rightarrow 0$ in its
$2J'$-dimensional expression (\ref{newangle}). As for $\tilde{Z}$, it is
therefore the $2J'\times2J'$ diagonal matrix obtained from $Z$ by
removing its top eigenvalue $J-1/2$:
\beq
\tilde{Z} & = & \mbox{Diag}(-J+1/2,-J+3/2,\cdots,J-3/2)=\mbox{Diag}
(-J',-J'+1,\cdots,J'-1)\\ \Rightarrow \tilde{Z}_{ij} & = & (-J'-1+i)
\delta_{i,j}.
\eeq
Note that $Z$ and $\tilde{Z}$ are defined by their commutation
relation (\ref{angle}) and only appear in the action through
$\partial_\phi$ as a commutator. As a result, they are only defined up
to a translation by a matrix proportional to the unit, and thus
$\tilde{Z}=\mbox{Diag}(1,\cdots,2J')$ is another possible choice.

Although $\widetilde{e^{i\phi}}$ is not unitary, the equation
\begin{eqnarray*}
<m'|[\tilde{Z},\widetilde{e^{i\phi}}]|m> & = & (m'-m)\delta_{m',m+1}=\delta 
_{m',m+1}=<m'|\widetilde{e^{i\phi}}|m>\mbox{ if } m<2J'\\
& = & 0=<m'|[\tilde{Z},\widetilde{e^{i\phi}}]|2J'>\mbox{ if }
m=2J'
\end{eqnarray*} 
shows that $\tilde{Z}$ and $\widetilde{e^{i\phi}}$ do satisfy the
commutation relation (\ref{angle}) for $\alpha=1$

\subsection{Expressing the Kinetic term}
Using the reduced action (\ref{reducedS}), the kinetic term $K(\Phi)$ then
takes the general form  
\be 
K(\Phi)=Tr\left(|[\tilde{Z},\Phi]~|^2~+~|[\widetilde{e^{i\phi}},\Phi]~|^2
\right)=\sum_{i,j=1}^{2J'} \left[ (i-j)^2|\Phi_{ij}|^2+|\Phi_{i-1\,j}-\Phi
_{i\,j+1}|^2\right] ,\label{Kphi}
\ee
where we have introduced new entries $\Phi_{0j}=\Phi_{i\,2J'+1}=0$ set
to zero to simplify the expressions.

The kinetic term can be further reordered as 
\beq
K(\Phi) & = & \sum_{i=1}^{2J'}\sum_{j=1}^{i-1}[2(i-j)^2+4-\delta_{i,N}-\delta_{j,1}] 
|\Phi_{ij}|^2+\sum_{i=1}^{2J}(2-\delta_{i,N}-\delta_{i,1})|\Phi_{ii}|^2
- \\ & & 4\sum_{i=1}^{2J'-1}\sum_{j=1}^{i-1}\mathcal{R}[\Phi_{ij}^*
\Phi_{i+1\,j+1}]-2\sum_{i=1}^{2J'-1}\Phi_{ii}\Phi_{i+1\,i+1}\nonumber
\eeq
Note in this expression that the last two sums are {\bf not} over all possible
indices, but omitting  the highest one. Furthermore, in the last two terms,
$\Phi_{ij}$ is coupled to both $\Phi_{i+1\,j+1}$ and $\Phi_{i-1\,j-1}$

The expression of the potential (\ref{potential}) is already known,
being the same expression as for the fuzzy sphere (see
e.g. \cite{xavier}).

\subsection{The action on a fuzzy cylinder of radius $r$}
The cylinder is also parametrized by its radius $r$. According to
(\ref{qntmalg}), $\hat{r}$ commutes with both $Z$ and
$e^{i\phi}$. It can therefore be considered as a pure number in the
non-commutative cylinder algebra.

The radius will appear as a simple scaling in the action. The volume
of the cylinder $Tr(1)$ depends linearly on $r$, so the action should
have an overall scale of $r$. The derivative along the axis
$\partial_Z$ does not scale with $r$, whereas the angular derivative
$\partial_\phi$ scales like $1/r$.

As a result, the action on a fuzzy cylinder of radius $r$ is given by:
\beq
S&=&r\,\widetilde{Tr}\left(\frac{1}{r^2}|~[Z,\Phi]~|^2 ~+~ |~e^{-i\phi}[e^{i\phi},\Phi]~|^2~+~V(\Phi)\right)\\
&=& r~Tr_{J'}\left(\frac{1}{r^2}|~[\tilde{Z},\Phi]~|^2~+|~[\widetilde{e^{i\phi}},
\Phi]~|^2+~V(\Phi)\right) \label{actionr}
\eeq

\section{Spectrum of the Laplacian on the fuzzy cylinder}
The Laplacian comes from the kinetic term (\ref{Kphi}) of the action. After
integrating by parts\footnote{Since commutators work as derivations
  and the trace as integration, there is a direct equivalent to integration
  by part given as $$Tr(\Phi
  [L,\Psi])=Tr([L,\Phi\Psi])-Tr([L,\Phi]\Psi)=-Tr([L,\Phi]\Psi)$$}, we get
$$K(\Phi)=Tr(-[\tilde{Z},\Phi][\tilde{Z},\Phi]-[\widetilde{e^{i\phi}}
^\dag,\Phi][\widetilde{e^{i\phi}},\Phi])=Tr(\Phi[\tilde{Z},[\tilde{Z},\Phi]]
+\Phi [\widetilde{e^{i\phi}}^\dag,[\widetilde{e^{i\phi}},\Phi]])$$
so that naively
$$\cL^2\Phi=[\tilde{Z},[\tilde{Z},\Phi]]+[\widetilde{e^{i\phi}}^\dag,[
\widetilde{e^{i\phi}},\Phi]]=\mathcal{L}^2_Z\Phi+\mathcal{L}_{_-}
\mathcal{L}_{_+}\Phi,$$
where $\mathcal{L}_{Z}$, resp.$\mathcal{L}_{_+}$, resp. $\mathcal{L}
_{_-}$, is the adjoint action of $Z$, resp. $\widetilde{e^{i\phi}}$,
resp. $\widetilde{e^{i\phi}}^\dag$. Note however that this 
Laplacian is {\em not} hermitian due to $\widetilde{e^{i\phi}}$ not
being actually unitary in the second term. To make it hermitian, the
latter term must be symmetrize. The Laplacian now is:
\beq
\cL^2\Phi & = & [\tilde{Z},[\tilde{Z},\Phi]]+\frac{1}{2}([\widetilde{e
^{i\phi}}^\dag,[\widetilde{e^{i\phi}},\Phi]]+[\widetilde{e
^{i\phi}},[\widetilde{e^{i\phi}}^\dag,\Phi]]) \nonumber \\ & = & \mathcal{L}^2_Z
\Phi+\frac{1}{2}(\mathcal{L}_{_-}\mathcal{L}_{_+}\Phi+\mathcal{L}_{_+}
\mathcal{L}_{_-}\Phi) \label{Laplacian} \\
& = & \mathcal{L}^2_Z\Phi+\mathcal{L}_{_-}\mathcal{L}_{_+}\Phi+
\frac{1}{2}[[\widetilde{e^{i\phi}},\widetilde{e^{i\phi}}^\dag],
\Phi]=\mathcal{L}^2_Z\Phi+\mathcal{L}_{_-}\mathcal{L}_{_+}\Phi+\frac{1}{2}
[\mbox{Diag}(-1,0,\cdots,0,1),\Phi]\nonumber\eeq

The Laplacian derived above in (\ref{Laplacian}) is for a cylinder of
radius one. The action (\ref{actionr}) on a cylinder of radius $r$  
shows how the Laplacian scales with the radius $r$, yielding
\be \cL^2\Phi=\frac{1}{r^2} \mathcal{L}^2_Z\Phi+\frac{1}{2}(
\mathcal{L}_{_-}\mathcal{L}_{_+}\Phi+\mathcal{L}_{_+}\mathcal{L}_{_-}\Phi).
\label{Laplacianr}\ee

Since we will be interested in the entropy of the free field, let us
now look for the eigenvalues of the Laplacian. 

\subsection{Symmetries of the Laplacian:} Because the
Laplacian is expected to be invariant with respect to $Z$-axis
rotations, it must commute with $\mathcal{L}_Z$. This is quite obvious
on the expression of the Laplacian since: 
\begin{itemize}
\item[-] $\widetilde{e^{i\phi}}$ has axial momentum $+1$ and will
therefore raise total axial momentum by one whether multiplied on
the left or on the right
\item[-] conversely $\widetilde{e^{i\phi}}^\dag$ has axial momentum
$-1$ and will therefore lower total axial momentum by one whether
multiplied on the left or on the right 
\end{itemize}
and therefore overall the axial angular momentum is conserved by
$\mathcal{L}_{_-}\mathcal{L}_{_+}$ or $\mathcal{L}_{_-}\mathcal{L}_{_+}$.
Taking into account that we have hermitian eigenmatrices, we can
deduce that they must be a mix of $+m$ and $-m$ axial momentum 
matrices. This means that we can take the ansatz 
\be
\Phi=\Phi_m(\vec{d})+\Phi_m^\dag(\vec{d}),\mbox{ with } 
\Phi_m(\vec{d})=\sum_{i=1}^{2J'-m} d_i\,|i><i+m|,\ 0\leq m\leq
2J'-1 \label{evansatz} 
\ee 
for the eigenmatrices, where the vector $\vec{d}=(d_i)_{1\leq i\leq
2J'-m}$ is the unknown to be determined by the eigenmatrix equations.

Using this ansatz (\ref{evansatz}) and the hermiticity of $\Phi$ and
the Laplacian, the eigenmatrix equation for the
Laplacian reads simply
$$\mathcal{L}^2\Phi_m(\vec{d})=\lambda\Phi_m(\vec{d})=\Phi_m(\lambda\vec{d})
\Leftrightarrow 
M_m\vec{d}=\lambda\vec{d}$$
where the matrix $M_m$ translates the (linear) action of the Laplacian
on $\Phi_m(\vec{d})$ to $\vec{d}$: $\mathcal{L}^2\Phi_m(\vec{d})=
\Phi_m(M_m\vec{d})$. This $2J'-m$ square matrix is now evaluated.

Note that the Laplacian has another less obvious symmetry, it is
invariant under the replacement $|i>\rightarrow |2J'-i>$. We will not
need to use it in the following calculation, but the eigenvector
equations in the following should be (and we checked that they
actually are) invariant under that symmetry. 

\subsection{The matrix $M_m$}
\paragraph*{\underline{Piece of $M_m$ coming from $\mathcal{L}_Z^2$:}}
In this case, since by  construction, $\Phi_m$ is an eigenmatrix of
$\mathcal{L}_Z$ with axial angular momentum $m$,
$$\mathcal{L}_Z^2\Phi_m(\vec{d})=m^2\Phi_m(\vec{d})=\Phi_m(m^2
\vec{d}).$$

\paragraph*{\underline{Piece of $M_m$ coming from $\mathcal{L}_{\{_+}
\mathcal{L}_{_-\}}$:}}
\begin{eqnarray*} 
\mathcal{L}_{\{_+}\mathcal{L}_{_-\}}\Phi_m(\vec{d}) & = & \sum_i d_i 
\mathcal{L}_{\{_+}\mathcal{L}_{_-\}}(|i><i+m|) \\
& = & \sum_i d_i \left( \{ x_+,x_-\}|i><i+m|+|i><i+m|\{ x_+,x_-\}-
\right. \\ & & \left. x_+|i><i+m|x_--x_-|i><i+m|x_+ \right) \\
& = & \sum_i d_i \left( (1-\delta_{i,1}/2-\delta_{i,2J'}/2+1-\delta
_{i+m,1}/2-\delta_{i+m,2J'}/2)|i><i+m|- \right. \\
& & (1-\delta_{i,2J'})(1-\delta_{i+m,2J'})|i+1><i+m+1|- \\ & & \left. 
(1-\delta_{i,1})(1-\delta_{i+m,1})|i-1><i+m-1| \right) \end{eqnarray*}
where $x_-=\widetilde{e^{i\phi}}^\dag$ lowers the index of the ket,
whereas $x_+=x_-^\dag$ raises it. At this point in the calculation, it
is necessary to distinguish the cases $m=0$, for which all the Kronecker can
be $0$, and $m\not=0$ where only half of them can.

For $m=0$, we find 
\begin{eqnarray*} 
\mathcal{L}_{\{_+}\mathcal{L}_{_-\}}\Phi_m(\vec{d}) & = &
\sum_i d_i \left( (2-\delta_{i,1}-\delta_{i,2J'})|i><i|-(1-
\delta_{i,2J'})|i+1><i+1|-\right. \\ && \left. (1-\delta_{i,1})
|i-1><i-1| \right) \\
& = & \Phi_m(\left(\begin{array}{cccccc} 1&-1&0&&\\-1&2&-1&&(0)\\
&\ddots&\ddots&\ddots&\\(0)&&\ddots&2&-1\\&&&-1&1
\end{array}\right)\vec{d})\end{eqnarray*}

For $m\not=0$, on the other hand,
\begin{eqnarray*} 
\mathcal{L}_{\{_+}\mathcal{L}_{_-\}}\Phi_m(\vec{d}) & = &
\sum_i d_i \left( (2-\delta_{i,1}/2-\delta_{i+m,2J'}/2)|i><i+m|- \right. \\
& & \left. (1-\delta_{i+m,2J'})|i+1><i+m+1|-(1-
\delta_{i,1})|i-1><i+m-1| \right) \\
& = & \Phi_m(\left(\begin{array}{cccccc} 3/2&-1&0&&\\-1&2&-1&&(0)\\
&\ddots&\ddots&\ddots&\\(0)&&\ddots&2&-1\\&&&-1&3/2\end{array}\right)
\vec{d}) \end{eqnarray*}

\paragraph*{\underline{Expression of $M_m$}} 
Putting together the results from the last two paragraph, and using
the Laplacian (\ref{Laplacianr}) on a cylinder of radius $r$, we get\be 
M_0=\left(\begin{array}{cccccc} 1&-1&0&&\\-1&2&-1&&(0)\\
&\ddots&\ddots&\ddots&\\(0)&&\ddots&2&-1\\&&&-1&1\end{array}\right),\ 
M_m=\frac{m^2}{r^2}+\left(\begin{array}{cccccc} 3/2&-1&0&&\\-1&2&-1&&(0)\\
&\ddots&\ddots&\ddots&\\(0)&&\ddots&2&-1\\&&&-1&3/2\end{array}\right),\ m
\not=0\ee
These matrices are similar to the ones obtained for the Laplacian on a
one-dimensional lattice and can actually be diagonalized without much difficulty,
taking good care to remember that $M_m$ is a matrix of dimension
$2J'-m$.

\subsection{Spectrum of $M_m$}
The way to evaluate the spectrum is to write explicitly the
eigenvector equation. Then, let $\vec{d}=(d_i)$ be an eigenvector for
the eigenvalue $\lambda$, the eigenvector equation takes the form
\be d_{n+1}=(2-\lambda)d_n-d_{n-1}, 2\leq n < 2J'-m \label{induction} \ee
plus boundary equations at each end $n=1,2J'-m$, which are different
for $m=0$ and $m\not=0$.

The sequence defined by this linear induction formula with constant
coefficients can be determined by
looking at its characteristic equation\footnote{The linear space of
  sequences satisfying the induction formula (\ref{induction}) is of
  dimension $2$, parametrized by the two initial values of the
  sequence. The idea is to find a basis of this linear space in the
  form of two geometric sequences of the form $(q^n)$. $q$ must then
  satisfy a (quadratic) characteristic equation, and the sequence
  we want to express is a linear combination of these two geometric
  sequences.}  
 $q^2-(2-\lambda)q+1=0$. Reparametrizing the eigenvalues as 
\be \lambda=2-2\cos(\theta)=4\sin^2(\t/2),\label{ev}\ee 
this equation has the simple solutions $q=\exp(\pm i\theta)$. Therefore, the
sequence has the general form \be
d_n=\cR(\b e^{in\theta}) \label{dn} \ee
where $\beta$ is a complex constant to be determined by the first two
terms of the sequence. 

To simplify the algebra, it is convenient to extrapolate $d_0$ and
$d_{2J-m+1}$ and rewrite the boundary equations between them, $d_1$, and
$d_{2J-m}$. Furthermore, since the eigenvector is defined up to an
overall constant, let us choose $d_0=1$. 

Now we must look at the two cases separately.

\paragraph*{\underline{Spectrum of $M_0$}}
In this case, the boundary equations can be seen to take the form $$
\left\{\begin{array}{ccl} d_2 & = & (1-\lambda)d_1 \\ d_2 & = &
(2-\lambda)d_1-d_0 \end{array}\right. \Rightarrow d_1=d_0.$$
Similarly,  at the other boundary,
\be d_{2J'-1}=d_{2J'}.\label{constr0}\ee

In particular, $d_0=d_1=1$. Plugging those initial values in the
general form (\ref{dn}) of the sequence, we get
$$\left\{\begin{array}{ccl} \cR(\b) & = & 1\\ \cR(\b e^{i\t}) & = & 1
\end{array}\right. \Leftrightarrow \b=1-i\tan(\t/2)=\frac{e^{-i\t/2}}{\cos(
\theta/2)},$$
and therefore 
\be d_n=\frac{1}{\cos(\theta/2)}\cos((n-1/2)\t).\label{dn0}\ee
To get the eigenvalues, it only remains to enforce the last constraint
equation (\ref{constr0}):
$$\cos((2J'-1/2)\t)=\cos((2J'+1/2)\t)\Leftrightarrow \sin(2J'\t)\sin(\t/2)=0
\Leftrightarrow \t= k\pi/2J',\ 0<k\leq 2J',$$
which, according to (\ref{ev}) gives eigenvalues: \be
\lambda_0^k=4\sin^2(k\pi/4J'),\ 0<k\leq 2J'.\label{ev0}\ee
A corresponding eigenvector is then given in (\ref{dn0}) up to an overall constant as:
\be \vec{d}=\left(\cos((n-1/2)k\pi/2J')\right)_{1\leq n\leq 2J'}.\ee

\paragraph*{\underline{Spectrum of $M_m$, $m\not=0$}}
In this case, the boundary equations can be seen to take the form 
$$\left\{\begin{array}{ccl} d_2 & = & (3/2-\lambda)d_1 \\ d_2 & = &
(2-\lambda)d_1-d_0 \end{array}\right. \Rightarrow d_1=2d_0 $$
And similarly, at the other boundary, \be
2d_{2J'+1-m}=d_{2J'-m}. \label{constrm}\ee
In particular, $d_0=1$ and $d_1=2$. Plugging those initial values in the
general form (\ref{dn}) of the sequence, we get
$$\left\{\begin{array}{ccl} \cR(\b) & = & 1\\ \cR(\b e^{i\t}) & = & 2
\end{array}\right. \Leftrightarrow \b=1+i\frac{\cos(\t)-2}{\sin(\t)},$$
and therefore 
\be d_n=\frac{1}{\sin(\t)}\mathcal{I}(2e^{in\t}-e^{i(n-1)\t})=\frac{2
\sin(n\theta)}{\sin(\theta)}-\frac{\sin((n-1)\theta)}{\sin(\theta)}.
\label{dnm}\ee
To get the eigenvalues, it only remains to enforce the last constraint
equation in (\ref{constrm}). Denoting $N=2J'-m$, and using the
trigonometric relation $$\sin(a)-\sin(b)=2\cos((a+b)/2)\sin((a-b)/2),$$
we get: 
\begin{eqnarray}
&& 2(\sin((N+1)\t)-\sin(N\t))=\sin(N\t)-\sin((N-1)\t)\nonumber \\ 
\Leftrightarrow &&
2\cos((N+1/2)\t)=\cos((N-1/2)\t)\Leftrightarrow 3\tan(N\t)\tan(\t/2)=1 
\nonumber \\
\Leftrightarrow && \tan(N\t)=\tan((\pi-\t)/2)/3.\label{evEq}
\end{eqnarray}
which can be seen graphically to have $N$ solutions, one in each
interval $[k\pi/N;(k+1)\pi/N ]$, $0\leq k< N$. These solutions
must be determined numerically though. The eigenvalues are then
deduced from Eq. (\ref{ev}). The corresponding eigenvector is then
given by (\ref{dnm}).

For large matrices $N\gg 1$, it is possible to find approximate
solutions since $\tan(x)$ can be well approximated near $0$ and
$\pi/2$. This yields: \begin{itemize} 
\item For $\theta\ll \pi$,  or $k\ll N$, $\tan((\pi-\t)/2)\sim 2/\t\gg
  1$. Therefore,  $N\t=k\pi+\pi/2-\rho_k$ with $\rho_k\ll 1$. The
  equation then becomes:
$$\frac{1}{\rho_k}\simeq \frac{2N}{3(k\pi+\pi/2-\rho_k)}\Leftrightarrow
\rho_k\simeq\frac{3\pi}{2N}(k+1/2).$$
\item For $\theta\simeq \pi$, or $k\sim N$,
  $\tan((\pi-\t)/2)\sim (\pi-\t)/2\ll 1$, and therefore,
  $N\t=k\pi+\rho_k$ with $\rho_k\ll 1$. The equation then becomes:
$$\rho_k\simeq\frac{N\pi-k\pi-\rho_k}{6N}\Leftrightarrow
\rho_k\simeq \frac{N-k}{6N}\pi$$
which is a small number, as expected, since $k\sim N$.
\end{itemize}

\section{Numerical Simulations and Results}
\subsection{The numerical scheme}
The model defined by the action Eqs.(\ref{actionr},\ref{potential})
which we want to simulate has three parameters $(\mu,c,r)$ plus the
matrix size $J$.  The goal is to explore the parameter space for
various phases of $\Phi$.  The simulations are carried out using the
"pseudo-heat bath" Monte-Carlo (MC) algorithm \cite{digal1,digal2} to
reduce the auto-correlation along the MC history.

The field should also be allowed to explore the whole phase space and
not remain trapped in local minima. To this end, an over-relaxation
method, first suggested in \cite{panero}, is also used. Let us
introduce $S_\Phi(\Phi_{ij})$ the dependence of the action on the
field entry $\Phi_{ij}$ when the field takes the value $\Phi$. It is a
fourth degree polynomial. Therefore the equation
$S_\Phi(\Phi_{ij})=S_\Phi(\Phi_{ij}=a)$, which has an obvious solution
$\Phi_{ij}=a$, can be factorized into a degree three polynomial which
{\it always} admits at least one real solution. The overrelaxation
method consists in replacing the field entry $\Phi_{ij}=a$ by one of
these real solutions, thereby moving the field in a different region
of the phase space.

A crosscheck is also used to verify that the field probability
distribution of our Monte-Carlo runs are consistent. Let us split the
terms in the action according to their scalings
$$S(\phi)=S_2(\phi)+S_4(\phi)\mbox{ with } S_i(\lambda\phi)=\lambda^i
S_i(\phi).$$
Then one can define a modified partition function 
\beq
Z(\lambda)=\int[\diff\phi]e^{-S(\lambda\phi)} & = & \int[\diff\phi]
e^{-\lambda^2S_2(\phi)-\lambda^4S_4(\phi)} \label{Z1} \\
& = & \lambda^{-N}\int[\diff\psi]e^{-S(\psi)},\ \psi=\lambda\phi,
\label{Z2} 
\eeq
where $N$ is the number of degrees of freedom in the field $\phi$
which appear in the integration. 
Evaluating  
\begin{eqnarray*}
\left.\frac{\partial \ln(Z)}{\partial\lambda}\right|_{\lambda=1} & = & 
-2<S_2>-4<S_4> \mbox{ from (\protect\ref{Z1})} \\ 
& = & -N\mbox{ from (\protect\ref{Z2})} 
\end{eqnarray*}
yields the check originally due to Denjoe O'Connor \cite{denjoe1}.
\be
<S_2>+2<S_4>=N/2
\label{Did}\ee
In all simulations, this identity (\ref{Did}) is always satisfied to better than $1\%$ relative error.

\subsection{The phase structure}
The temperature ($T$) is regulated by varying the parameter
$\mu$. \begin{itemize} 
\item $\mu\ll 1$ corresponds to low temperatures when the fluctuations are small. In
this case, the minimum of $S$ gives the most probable configuration of
the phase. In Eq. (\ref{actionr}), it is possible to minimize
the action by minimizing separately the kinetic term, so that $\Phi
\propto {\bf 1}$, and the potential term so that
$\Phi=\sqrt{-\mu/2c}\,{\bf 1}$, and this phase is therefore known as the
uniform phase. 
\item At high temperatures, $ \mu\gg 1$, the thermal
fluctuations lead the system to the disorder phase $\Phi \sim 0$. 
\item At intermediate temperatures, the competition between the action and the
fluctuations give rise to new phases called the non-uniform or stripe
phases. These new phases are specific to non-commutative
spaces. Various numerical studies have confirmed the existence of
these phases \cite{xavier,denjoe,panero,digal1,digal2,bietenholz,
  medina,ambjorn} on the fuzzy sphere. 
A non-commutative cylinder will also exhibit the non-uniform
phases. However, due to the non-trivial topology of the cylinder (the
first homotopy group being non-trivial), one can have a more complex
phase structure described below.
\end{itemize}

For example there can be stripes going 
around the cylinder, or parallel to its axis. These two phases can be 
distinguished by their overlap with the operators $Z$, ${e^{i\phi}}$,
and ${e^{i\phi}}^\dag$   
respectively. Stripes going around the cylinder will have non-zero 
overlap with the operator $Z$. While a configuration of stripes along 
the axis will have overlap with ${e^{i\phi}}$ and
${e^{i\phi}}^\dag$. We present our results in the following subsection.

\subsection{Example numerical runs}
For a given choice of $N=7$, $c = 0.36$, and $r=1$ the simulations are
done for various values of $\mu$. The various phases discussed above can
be characterized by the observables $m_u=Tr(\Phi)$, $m_z=Tr(\Phi Z)$,
$m_x = Tr(\Phi e^{i\phi})$. A finite $m_u$ with $(m_z,m_x) \sim 0$
characterizes the uniform phase. On the other hand, $(m_u,m_x)\sim 0$
with non-zero $m_z$ characterizes stripes going around the
cylinder. Stripes along the cylinder characterized by $(m_u,m_z)\sim
0$ with non-zero $m_x$.

\begin{figure}[hbt!]
\begin{center}
{\includegraphics[scale=0.79]{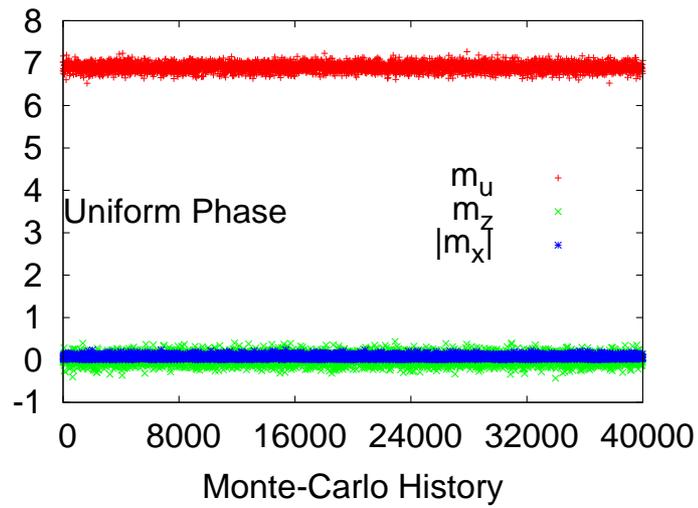}}
\caption{$m_u,m_z,m_x$ vs Monte Carlo history for $N=7$, $\mu = -35.1$,
  $c = 0.36$, and $r=1$.} 
\label{uprun}
\end{center}
\end{figure}

For $\mu = -35.1$, the data of a run are shown on Figure  \ref{uprun},
and, as expected, we observe the uniform phase. 

\begin{figure}[hbt!]
\begin{center}
{\includegraphics[scale=0.79]{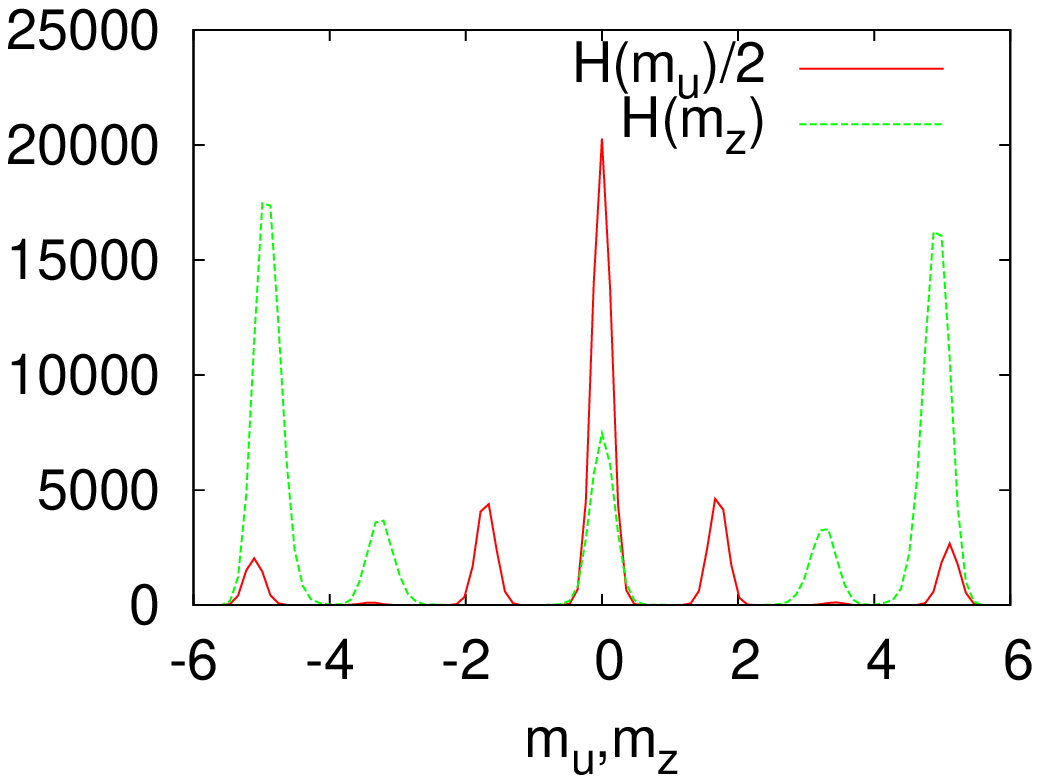}}
\caption{Histogram of $H(m_u)$ and $H(m_z)$ $N=7$, $\mu = -20.0$,
  $c = 0.36$, and $r=1$.}
\label{histo}
\end{center}
\end{figure}

For $ \mu = -20.0$, we observed the phase with stripes going around the
cylinder. This is verified on the histogram of the observed values of
$m_u,m_z$ plotted in Figure \ref{histo}. It is clear from the figure
that the average value of $m_z$ is finite while the average value of
$m_u$ is vanishingly small.

\begin{figure}[hbt!]
\begin{center}
{\includegraphics[scale=0.79]{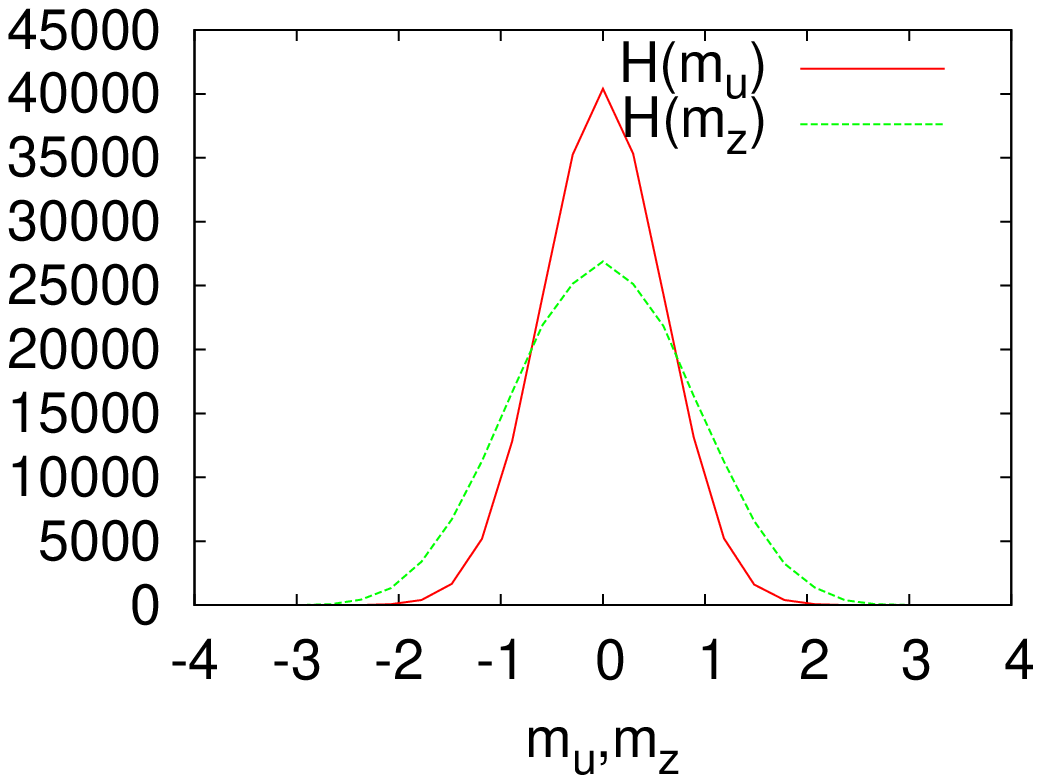}}
\caption{Histogram of $H(m_u)$ and $H(m_z)$ $N=7$, 
  $c = 0.36$, and $r=1$.}
\label{disrun}
\end{center}
\end{figure}

Figure \ref{disrun} shows the system in the disorder phase where
$m_u,m_z,m_x$ all fluctuate around zero. 

\begin{figure}[hbt!]
\begin{center}
{\includegraphics[scale=0.79]{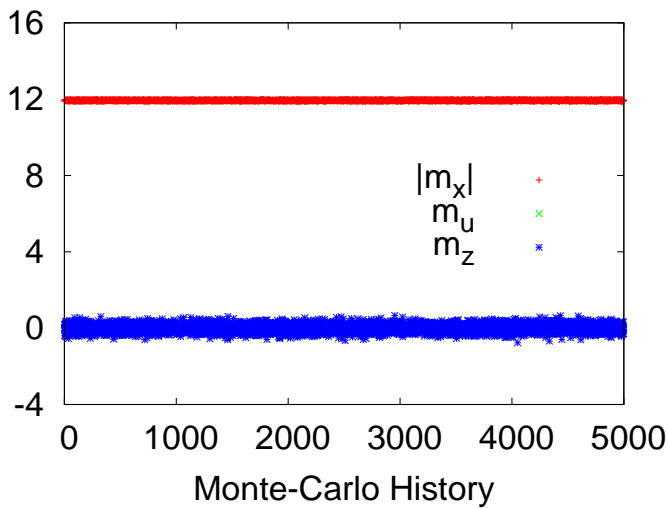}}
\caption{$m_u,m_z,m_x$ vs Monte Carlo history for $N=7$, $\mu= -3.6
  \times 10^3$, $c=36.$, $r=0.01$.}
\label{stripesrun}
\end{center}
\end{figure}

We did not observe the phase with stripes going along the cylinder as
a ground state for any choice of $\mu$ for $r\sim 1$. One can expect
to observe this state for very small $r$ when the second term of the
kinetic term in the action (\ref{actionr}), which suppresses this state,
is made subdominant. For a very small radius $r=0.01$, $c=36.$ and
$\mu= -3.6 \times 10^3$, this phase appears as meta-stable in Figure
\ref{stripesrun}. This phase is stable w.r.t small fluctuations. Only
large fluctuations, which occur less frequently, can destroy such a
state.

\newpage

\newpage

\noindent

\section{Conclusions}
In this paper, we have considered a finite dimensional representation of
the noncommutative cylinder algebra, which makes it fuzzy. We study
scalar field theory in the background of this algebra both analytically
and using numerical simulations.

The action of the scalar field on the fuzzy cylinder contains a
kinetic as well as a potential term. The kinetic term leads to the
Laplacian (\ref{Laplacianr}) on the fuzzy cylinder. We have analyzed
the symmetries of the Laplacian and have obtained an algebraic
equation (\ref{ev0},\ref{evEq}) describing the corresponding spectrum.

In the numerical simulations of the scalar field with a generic potential 
we find, as expected in noncommutative theories, novel stripe phases
breaking rotational symmetry.
But they have some differences with the usual stripes on 
Moyal spacetimes. These are also stable due to topological features arising 
in this fuzzy geometry. A similar stability has been 
seen in the $O(3)$ model on fuzzy spheres \cite{digaltrg}.

The fuzzy cylinder algebra considered in this paper is valid for non-extremal BTZ black holes, without any further assumptions.  In addition, different forms of the BTZ metric related by coordinate transformations are equivalent
classically, which at the algebraic level are expected to be related by automorphisms.

While the noncommutative cylinder algebra first arose in the
context of the BTZ black hole, subsequently it has been shown to be of
more general relevance, appearing in diverse backgrounds such as for Kerr
black hole \cite{schupp} and FRW cosmologies \cite{ohl}. In addition, the near-horizon
geometry of a large class of black holes contains an $AdS_3$ factor \cite{ooguri}. 

Thus, a large class of noncommutative black holes are
described by a noncommutative cylinder algebra. The fuzzy cylinder algebra
derived from it can therefore be used to define a fuzzy black hole. From
general considerations \cite{dop2}, we know that such
black holes can arise at the Planck scale. Our results provide a first
glimpse about the phase structure of a quantum scalar field theory in the
background of a fuzzy black hole at the Planck scale.

{\bf Acknowledgment} This work was done as a part of the CEFIPRA/IFCPAR project 4004-1 entitled Fuzzy Approach to Quantum Field Theory and Gravity. The authors gratefully acknowledge the financial assistance from CEFIPRA/IFCPAR which was essential for this work.


\begin{thebibliography}{99}
 

\bibitem{dop2} S. Doplicher, K. Fredenhagen and J. E. Roberts, Commun. Math. Phys. {\bf 172} (1995) 187.

\bibitem{wess1} P. Aschieri, C. Blohmann, M. Dimitrijevic, F. Meyer, P. Schupp and J. Wess, Class. Quant. 
Grav. {\bf 22} (2005) 3511.


\bibitem{seckin} A.P. Balachandran, T.R. Govindarajan, K.S. Gupta, S. Kurkcuoglu,  Class. Quant. Grav. {\bf 23} (2006) 5799.

\bibitem{brian1} B.P. Dolan, Kumar S. Gupta and A. Stern, Class. Quant. Grav. {\bf 24} (2007) 1647. 


\bibitem{schupp} P. Schupp and S. Solodukhin, arXiv:0906.2724 [hep-th].

\bibitem{ohl} T. Ohl and A. Schenkel, JHEP {\bf 0910} (2009) 052. 

\bibitem{btz1} M. Banados, C. Teitelboim and J. Zanelli, Phys. Rev. Lett. {\bf 69} (1992) 1849.

\bibitem{btz2} M. Banados, M. Henneaux, C. Teitelboim and J. Zanelli, Phys. Rev. {\bf D  48} (1993) 1506.
\bibitem{petergrosse} H. Grosse and P. Presnajder, Acta Phys. Slovaca {\bf 49} (1999) 185.

\bibitem{L1} J. Lukierski, A. Nowicki, H. Ruegg and V. N. Tolstoy, Phys. Lett.  {\bf B 264} (1991) 331.



\bibitem{L4} J. Lukierski, H. Ruegg and W. J. Zakrzewski, Ann. Phys. {\bf 243} (1995) 90. 

\bibitem{M1} S. Meljanac and M. Stoji\'{c}, Eur. Phys. J. C \textbf{47} (2006) 531.


\bibitem{M3} S. Meljanac, A. Samsarov, M. Stoji\'{c} and K. Gupta, Eur. Phys. J. C \textbf{53} (2008) 295.


\bibitem{luk2} M. Daszkiewicz, J. Lukierski and M. Woronowicz, J. Phys. {\bf A 42} (2009) 355201. 

\bibitem{luk3} J. Lukierski, Rept. Math. Phys. {\bf 64} (2009) 299.

\bibitem{G1} T.~R.~Govindarajan, K.~S.~Gupta, E.~Harikumar, S.~Meljanac and D.~Meljanac, Phys.\ Rev.\  D {\bf 77} (2008) 105010.

\bibitem{G2} T.~R.~Govindarajan, K.~S.~Gupta, E.~Harikumar, S.~Meljanac and D.~Meljanac, Phys.\ Rev.\  D {\bf 80} (2009) 025014.

\bibitem{dice2010} T.R.Govindarajan, Kumar S. Gupta, E. Harikumar and S. Meljanac, Journal of Physics: Conference Series 306 (2011) 012019.

\bibitem{ksms} Kumar S. Gupta, S. Meljanac and A. Samsarov, arXiv:1108.0341 [hep-th]

\bibitem{C2} M. Chaichian, A. Demichev, P. Presnajder and A. Tureanu, Phys. Lett. {\bf B 515} (2001) 426.

\bibitem{paulo} A. P. Balachandran, T. R. Govindarajan, A. G. Martins and P. Teotonio-Sobrinho, JHEP {\bf 0411} (2004) 068.

\bibitem{madore} J. Madore, {\it An Introduction to Noncommutative Differential Geometry and its Physical Applications}, (London Mathematical Society Lecture Note Series).

\bibitem{balbook} A.P. Balachandran, S. Kurkcuoglu and S. Vaidya, {\it Lectures on fuzzy and fuzzy SUSY physics}, World Scientific
(2007).

\bibitem{hoppe} J. Hoppe, Ph.D. Thesis, MIT (Cambridge MA, 1982).
\bibitem{pinzul} A.P. Balachandran, T.R. Govindarajan and B. Ydri,
Mod. Phys. Lett. {\bf A15} (2000) 1279 [hep-th/9911087];\\
A.P. Balachandran, A. Pinzul and B.A. Qureshi, JHEP
{\bf 0512} (2005) 002 [hep-th/0506037].

\bibitem{gubser} S.S. Gubser and S.L. Sondhi, Nucl. Phys. {\bf B605},
395 (2001) [hep-th/0006119].


\bibitem{xavier} X. Martin, JHEP {\bf 0404} (2004) 077 [hep-th/0402230].



\bibitem{denjoe} J. Medina, W. Bietenholz, F. Hofheinz and
D. O'Connor, PoS {\bf LAT2005} (2005) 263 [hep-lat/0509162];\\ 
F. G. Flores, D. O'Connor and X. Martin, PoS {\bf LAT2005} (2006) 262 [hep-lat/0601012];\\ 
D. O'Connor and B. Ydri, JHEP {\bf 0611} (2006) 016 [hep-lat/0606013];\\
J. Medina,  Phd. thesis, arXiv: 0801.1284 [hep-th]. 
\bibitem{bietenholz} W. Bietenholz, F. Hofheinz and J. Nishimura, 
Acta. Phys. Polon. {\bf B34} (2003) 4711 [hep-th/0309216];\\
Nucl. Phys. Proc. Suppl. {\bf 129} (2004)  865  [hep-th/0309182].
\bibitem{panero} M. Panero, SIGMA {\bf 2} (2006) 081 [hep-th/0609205];\\ 
JHEP {\bf 0705}, 082 (2007) [hep-th/0608202].

\bibitem{digal1} C.R. Das, S. Digal and T.R. Govindarajan, Mod. Phys. Lett. {\bf A23} 
(2008) 1781.
\bibitem{digal2} C.R. Das, S. Digal and T.R. Govindarajan, 
Mod. Phys. Letts {\bf A24} (2009) 2693.


\bibitem{ambjorn} J. Ambjorn and S. Catterall, Phys. Lett. {\bf B549} (2002) 
253 [hep-lat/0209106]. 

\bibitem{medina} J. Medina, W. Bietenholz and D. O'Connor,
JHEP {\bf 0804} (2008) 041  arXiv:0712.3366 [hep-th].

\bibitem{denjoe1} Denjoe O'Connor (private communication)

\bibitem{digaltrg} S. Digal and T. R. Govindarajan, arXiv:1108.3320 [hep-th]

\bibitem{ooguri} O. Aharony, S. S. Gubser, J. M. Maldacena, H. Ooguri and Y. Oz, Phys. Rept. {\bf 323} (2000) 183 arXiv:hep-th/9905111.




\end{thebibliography}
\end{document}